\documentstyle[11pt]{article}
\title{A Simpler Eulerian Variational Principle \\ for Barotropic
Fluids}
\author{Asher Yahalom \\ Faculty of Engineering, Tel-Aviv University\\
P.O.B. 39040, Ramat Aviv\\
 Tel-Aviv 69978, Israel\\
e-mail: asya@post.tau.ac.il}
\begin{document}
\maketitle

\newcommand{\beq}{\begin{equation}}
\newcommand{\enq}{\end{equation}}
\newcommand{\ber}{\begin{eqnarray}}
\newcommand{\enr}{\end{eqnarray}}
\newcommand{\nsc}{necessary and sufficient conditions }
\newcommand{\nec}{necessary }
\newcommand{\su}{sufficient }
\newcommand{\dl}{dimensionless }
\newcommand{\st}{stability }
\newcommand{\sa}{stationary }
\newcommand{\eq}{equation }
\newcommand{\eqs}{equations }
\newcommand{\br}{barotropic }
\newcommand{\er}[1]{\eq (\ref{#1}) }
\newcommand{\dr}[1]{definition (\ref{#1}) }

\begin{abstract}

The variational principle of barotropic Eulerian fluid dynamics
is known to be quite cumbersome containing as much as
eleven independent functions. This is much more than the
the four functions (density and velocity) appearing in the
Eulerian equations of motion. This fact may have discouraged
applications of the variational method. In this paper
a four function Eulerian variational principle is suggested and
the implications are discussed briefly.

\end{abstract}

\section{Introduction}

The motion of a fluid is usually described in two ways.
In the Lagrangian approach we follow the trajectory of each
fluid particle. While in the Eulerian approach we study the
evolution of the velocity and density fields.

Variational principles have been developed in both cases.
A variational principle in terms of Lagrangian variables
has been described by Eckart \cite{Eckart} and Bretherton \cite{Bretherton}.
While a variational principle in term of the Eulerian
variables is given by Herivel \cite{Herivel}, Serrin \cite{Serrin}, 
Lin \cite{Lin}
and Seliger \& Whitham \cite{Seliger}. 

The Eulerian approach appears to be 
much more appealing involving measurable quantities such as 
velocity and density fields instead of the trajectories of unseen particles
of the Lagrangian approach. 

However, regretfully the variational
principle of the Eulerian flow appears to be much more cumbersome than 
the Lagrangian case, containing
quite a few "Lagrange multipliers" and "potentials".
In fact the total number of independent functions in this 
formulation  according to the approach suggested by
Herivel \cite{Herivel}, Serrin \cite{Serrin} and Lin \cite{Lin}
is eleven which exceeds by many the four functions
of velocity and density appearing in the Eulerian equations of
a barotropic flow.

The variational principle of the Lagrangian approach
is on the other hand simple and straightforward to implement.
Bretherton \cite{Bretherton} has suggested to use the Lagrangian
variational principle with Eulerian variables and
especially constructed "Eulerian displacements" for which
the equations of motion are derived. This procedure appears to be
inconvenient for stability and numerical calculations.

Instead I intend to develop in this paper a simpler variational
principle in terms of Eulerian variables. This will be done by 
rearranging the terms in the original variational principle
 and using a simple trick to be discussed below.
The result will be a variational principle in terms of four 
Eulerian independent functions which is the appropriate number.  

The plan of this paper is as follows: first I review the 
original Eulerian variational principle and the equations 
derived from it. Than I derive a new variational principle
and discuss its implications.

\section{The Eulerian Approach to the Barotropic Flow}
\label{EAB}

In the Eulerian description of barotropic fluid dynamics we
seek a solution for the  velocity $\vec v = \vec v (x^k,t)$
and density
$ \rho = \rho (x^k,t)$ fields of a given flow. Those fields
depend on the space coordinates $ {\vec r} =(x^k)= (x,y,z)
~~[k,l,m,n = 1,2,3]$ and on the time variable $t$.
The fields are obtained by solving the Euler equations:
\beq
{\partial \vec v \over \partial t}  +
\vec v \cdot \vec \nabla \vec v + \vec \nabla (h + \Phi) = 0
\label{Eul}
\enq
in which $\vec \nabla= \frac{\partial}{\partial x^k}$.
The potential $ \Phi = \Phi (x^k,t)$ is a given function of
coordinates, while $h = h(\rho)$ is the specific enthalpy which
is a given function of $\rho$
usually designated as the equation of state.
In addition the density and velocity fields must satisfy
the continuity \eq:
 \beq
{\partial \rho \over \partial t}  +
\vec \nabla \cdot (\rho \vec v) = 0.
\label{Cont}
\enq
Thus we have to solve four \eqs in order to obtain four unknown
fields. This can be done when supplemented the appropriate
boundary conditions.

\subsection{The Classical Eulerian Variational Principle}
\label{OVP}

Researchers (Herivel \cite{Herivel}, Serrin \cite{Serrin}, Lin \cite{Lin})
seeking a variational principle
from which the Euler and continuity \eqs (\er{Eul} and \eq (\ref{Cont}))
can be derived, arrived at the action $A$
given in terms of the Lagrangian $L$ as:
\begin{eqnarray}
 A & = & \int ^{t_1}_{t_0} L dt \quad with \nonumber \\
\quad L & = & \int _{V} \left[{1 \over 2}
   \vec v^{2} - \varepsilon(\rho) -\Phi \right] \rho  d^{3}x \nonumber \\
   & + &
\int_{V} \left[ \nu \left({\partial \rho \over \partial t}  +
\vec \nabla \cdot (\rho \vec v)\right) -
\rho \vec \alpha \cdot \frac {D \vec \beta}{Dt}
 \right] d^3 x.
\label{Serrinact}
\end{eqnarray}
In which $\varepsilon(\rho)$ is the specific internal energy
connected to the specific enthalpy by the \eq:
\beq
h =  {\partial (\rho \varepsilon) \over \partial \rho}.
\label{hepsi}
\enq
The volume element is given by $d^3 x$ and the operator
$ \frac{D}{Dt} $ is defined such that:
$ \frac{D}{Dt} \equiv \frac{\partial}{\partial t}  +
\vec v \cdot \vec \nabla $. The variation principle contains in
addition to the desired four functions $\rho,\vec v$ also the
seven "potentials" $\nu,\vec \alpha, \vec \beta$ reaching to a total
of eleven variational variables. Varying the above action with
respect to those variables such that the variations vanish for
$\vec r$ and $t$ sufficiently large. We see that in order to
have $\delta A$ vanish for otherwise arbitrary $(\delta \vec v, \delta \rho,
\delta \nu,\delta \vec \alpha, \delta \vec \beta)$,
the following \eqs must be satisfied:
\ber
\frac {\partial \rho }{\partial t}  & + &
\vec \nabla \cdot (\rho \vec v) = 0
\label{ContinS}
\\
\frac {\partial (\rho \alpha_i)}{\partial t}  & + &
\vec \nabla \cdot (\rho \vec v \alpha_i) = 0
\label{alphaleq}
\\
\vec v  & = & \vec \alpha \cdot \vec \nabla \vec \beta + \vec \nabla \nu
\label{clebschl}
 \\
\frac {D \vec \beta}{Dt}  & = & 0
\label{betacon}
 \\
\frac {D \nu }{Dt} & = & \frac{1}{2} \vec v^2 - h - \Phi.
\label{nueq}
\enr
Combining \er{alphaleq} and \er{ContinS} we arrive at the
simpler \eq:
\beq
\frac {D \vec \alpha}{Dt} = 0.
\label{alphacon}
\enq
Calculating the expression $\frac {D \vec v}{Dt}$ using
\eqs (\ref{clebschl},\ref{betacon},\ref{alphacon},\ref{nueq})
we arrive at the \eqs of Euler:
\beq
\frac {D \vec v}{Dt}= {\partial \vec v \over \partial t}  +
\vec v \cdot \vec \nabla \vec v = -\vec \nabla (h + \Phi).
\label{Euls}
\enq
Thus the functions $\vec v$ and $\rho$ are extrema of the action
$A$ if they satisfy the Euler and continuity \eqs.

\subsection{The Variational Principle of Seliger \& Witham}
\label{VPSW}

Seliger \& Witham \cite{Seliger}  have proposed  to take
$\vec \alpha = (\alpha,0,0)$ and $\vec \beta = (\beta,0,0)$,
in this way one
obtains a variational principle with only seven functions.
Rewriting \eqs 
(\ref{ContinS},\ref{clebschl},\ref{betacon},\ref{nueq},\ref{alphacon})
for this case we obtain the following set of \eqs:
\ber
\frac {\partial \rho }{\partial t}  & + &
\vec \nabla \cdot (\rho \vec v) = 0
\label{ContinS2}
\\
\vec v  & = & \alpha \vec \nabla \beta + \vec \nabla \nu
\label{clebsch}
 \\
 \frac {D \alpha}{Dt} & = & 0
\label{alphaleq2}
\\
\frac {D \beta}{Dt}  & = & 0
\label{betacon2}
 \\
\frac {D \nu }{Dt} & = & \frac{1}{2} \vec v^2 - h - \Phi.
\label{nueq2}
\enr
We see that \eqs (\ref{ContinS2}, \ref{nueq2}) remain
unchanged. The velocity $\vec v$ is given now by the
Clebsch representation
(\er{clebsch}) and the entire set of \eqs are designated as
Clebsch's transformed \eqs of hydrodynamics (Eckart \cite{Eckart},
Lamb \cite{Lamb H.} p. 248). Bretherton \cite{Bretherton} have quoted a remark
by H. K. Moffat concerning the possibility of the function $\nu$
being not single valued. According to Moffat \cite{Moffat H.K.} the
helicity integral:
\beq
{\cal H} = \int \vec \nabla \times \vec v \cdot \vec v d^3 x
\label{helicity}
\enq
which is a measure of the knottedness of the vortex lines, must
be zero if $\nu$ is single valued which is clearly not true in
general. This can be shown as follows: suppose that $\nu$
is single valued than by inserting \er{clebsch} into \er{helicity}
and integrating by parts we obtain:
\beq
{\cal H} = \int \vec \nabla \nu \cdot \vec \nabla \alpha
\times \vec \nabla \beta  d^3 x =
\int \vec \nabla \cdot (\nu  \vec \nabla \alpha
\times \vec \nabla \beta ) d^3 x =
\int \nu \vec \nabla \times \vec v \cdot d \vec S.
\label{helicity2}
\enq
If we choose a volume made of closed vortex filaments such
that $\vec \nabla \times \vec v \cdot d \vec S = 0$
than clearly ${\cal H} = 0$. However, as Bretherton \cite{Bretherton}
noticed the \eqs of motion being {\it local} are unaffected
by global properties such as the non single-veluedness of $\nu$.

An analogue from classical mechanics may make things even
clearer. A few will object to describe the two dimensional
motion of a particle  moving under the influence of
the potential $V(R)$ (where $R$ is the radial coordinate),
by the coordinates $R,\phi$. Where $\phi$ is the azimuthal angel
which is not single valued. This form is found to be more
convenient than the single valued Cartesian $x,y$
representation.

\section{The Reduced Variational Principle }
\label{RVP}

Although Seliger \& Witham (1968) have managed to reduce the
Eulerian variational principle from eleven to seven functions.
The number of variational variables is still too much, since the
Eulerian \eqs contain only four unknown functions.
I intend to suggest a solution to this problem here.

First let us rewrite the Lagrangian of Seliger \& Witham
(\er{Serrinact} with $\alpha$ and $\beta$ being scalars):
\ber
L & = & \int_{V} \left[{1 \over 2}
   \vec v^{2} - \varepsilon(\rho) -\Phi \right] \rho  d^{3}x \nonumber \\
   & + &
\int \left[ \nu \left({\partial \rho \over \partial t}  +
\vec \nabla \cdot (\rho \vec v)\right) -
\rho \alpha  (\frac{\partial \beta}{\partial t} +
\vec v \cdot \vec \nabla  \beta)
 \right] d^3 x.
\label{Selwit}
\enr
Next we rearrange terms:
\ber
L & = & \int_{V} \left[- \rho \alpha \frac{\partial
\beta}{\partial t} + \nu \frac{\partial \rho}{\partial t} \right] d^{3}x
\nonumber \\
 & + & \int_{V} \left[{1 \over 2} \rho \vec v^{2} + \nu
 \vec \nabla \cdot (\rho \vec v) -
 \rho \alpha \vec v \cdot \vec \nabla \beta \right] d^{3}x
 - \int_{V} \left[\varepsilon(\rho) +\Phi \right] \rho d^{3}x.
\label{Selwit2}
\enr
Furthermore, we introduce the identities:
\ber
\nu \frac{\partial \rho}{\partial t} & = &
\frac{\partial(\nu \rho)}{\partial t} -
\rho \frac{\partial\nu }{\partial t} \nonumber \\
\nu \vec \nabla \cdot (\rho \vec v) & = &
\vec \nabla \cdot (\nu \rho \vec v) -
\rho \vec v \cdot \vec \nabla \nu.
\label{ident}
\enr
Inserting the above identities into \er{Selwit2} and rearranging
terms again we have:
\ber
L & = & -\int_{V} \left[\alpha \frac{\partial
\beta}{\partial t} + \frac{\partial \nu}{\partial t} \right]
\rho d^{3}x
\nonumber \\
 & + & \int_{V} \left[{1 \over 2} \rho \vec v^{2} -
 \rho \vec v \cdot (\alpha \vec \nabla \beta + \vec \nabla \nu)
\right] d^{3}x
- \int_{V} \left[\varepsilon(\rho) +\Phi \right] \rho d^{3}x
\nonumber \\
& + & \int_{V} \frac{\partial(\nu \rho)}{\partial t} d^{3}x +
\int_{V} \vec \nabla \cdot (\nu \rho \vec v) d^{3}x.
\label{Selwit3}
\enr
Now since:
\beq
{1 \over 2} \vec v^{2} -
\vec v \cdot (\alpha \vec \nabla \beta + \vec \nabla \nu)
= {1 \over 2} (\vec v  - \alpha \vec \nabla \beta -\vec \nabla
\nu)^2 - {1 \over 2} (\alpha \vec \nabla \beta +\vec \nabla
\nu)^2
\enq
We finally obtain:
\beq
L = L_r + L_v
\enq
in which:
\beq
L_r = \int_{V} \left[- (\alpha \frac{\partial
\beta}{\partial t} + \frac{\partial \nu}{\partial t})
 - {1 \over 2} (\alpha \vec \nabla \beta + \vec \nabla
\nu)^2 - \varepsilon(\rho) -\Phi \right] \rho d^{3}x
+ \frac{\partial \int_{V} \nu \rho d^{3}x}{\partial t}.
\label{Lr}
\enq
And
\beq
L_v = \int_{V} \frac{1}{2}(\alpha \vec \nabla \beta + \vec \nabla
\nu - \vec v)^2 \rho d^{3}x +
\int_{V} \vec \nabla \cdot (\nu \rho \vec v) d^{3}x.
\label{Lv}
\enq
The Lagrangian $L$ is dissected into a part depending on the
fluid velocity $\vec v$ which is denoted as $L_v$ and the
remaining part which does not depend on $\vec v$ which
is denoted $L_r$. Let us look at the action:
\beq
A_r = \int ^{t_1}_{t_0} L_r dt.
\enq
This action is a functional of the four variables:
$\alpha,\beta,\nu,\rho$. Note that the last term of $L_r$
is a full time differential and thus does not contribute to the \eqs
of motion. In order to simplify the variational calculations
we introduce the notations:
\beq
\vec u \equiv \alpha \vec \nabla \beta + \vec \nabla \nu
\quad {\rm and } \quad
\frac{\tilde D}{Dt} \equiv \frac{\partial}{\partial t}  +
\vec u \cdot \vec \nabla.
\label{udef}
\enq
Using the above notations and assuming that the arbitrary variations
$(\delta \rho, \delta \nu,\delta \alpha, \delta \beta)$ vanish for
$\vec r$ and $t$ sufficiently large we obtain that $\delta A_r = 0$
only if the following \eqs are satisfied:
\ber
\frac {\partial \rho }{\partial t}  & + &
\vec \nabla \cdot (\rho \vec u) = 0
\label{ContinSu}
\\
\frac {\partial (\rho \alpha)}{\partial t}  & + &
\vec \nabla \cdot (\rho \vec u \alpha) = 0
\label{alphalequ}
\\
\frac {\tilde D \beta}{Dt}  & = & 0
\label{betaconu}
 \\
\frac {\tilde D \nu }{Dt} & = & \frac{1}{2} \vec u^2 - h - \Phi.
\label{nuequ}
\enr
Combining\footnote{Compare the transition from \er{alphaleq} to
\er{alphacon}.}
\er{ContinSu} and \er{alphalequ} we arrive at the
following set of \eqs:
\ber
\frac{\partial \rho }{\partial t}  & + &
\vec \nabla \cdot (\rho \vec u) = 0
\label{Continu}
\\
\frac {\tilde D \alpha}{Dt}  & = & 0
\label{alphu}
\\
\frac {\tilde D \beta}{Dt}  & = & 0
\label{betu}
 \\
\frac {\tilde D \nu }{Dt} & = & \frac{1}{2} \vec u^2 - h - \Phi.
\label{nuu}
\enr
If we take $\vec u$ to be a Clebsch representation of some
velocity field $\vec v=\vec u$, than the above \eqs become the
Clebsch transformed \eqs of fluid motion (\eqs
\ref{ContinS2},\ref{alphaleq2},\ref{betacon2},\ref{nueq2}).
Thus we have achieved
a variational principle that does not contain $\vec v$ as a
{\it variational variable}. This situation in which only the
potentials appear in the variational principle but not the
physical velocity field $\vec v$ itself is familiar from other
branches of physics and will be discussed below.

It remains to discuss the Lagrangian $L_v$
given in \er{Lv}, using the theorem of
Gauss and the \dr{udef} of $\vec u$ we arrive at the form:
\beq
L_v = \int_{V} \frac{1}{2}(\vec u - \vec v)^2 \rho d^{3}x +
\int \nu \rho \vec v \cdot  d \vec S.
\label{Lv1}
\enq
If we take $\vec u$ to be a Clebsch representation of $\vec v$
than the first part of $L_v$ vanishes. Further more the surface
integral remaining vanish for both fluids contained in a vessel
in which $\vec v$ is parallel to the vessels surface and for
astrophysical flows in which the density vanishes on the
surface of the flow. However, in the case of helical flows
$\nu$ is not single-valued and one must add in addition to the
physical surface a "cut". Thus one obtains:
\beq
L_v = \int_{cut} [\nu] \rho \vec v \cdot  d \vec S
\label{Lv2}
\enq
where $[\nu]$ represents the discontinuity
of $\nu$ across the "cut". We conclude
that for helical flows $L_v$ does not vanish altogether.
And thus for those flows the new Lagrangian $L_r$ {\it is not}
equal to the Seliger \& Witham Lagrangian $L$,
the difference being a surface integral over the "cut".

\subsection{ A Comment Regarding Esthetics}

It was already noted that the last term of $L_r$ given in
\er{Lr} does not contribute to \eqs of motion. It seem esthetically
plausible to omit this term altogether thus obtaining:
\beq
L_s = \int_{V} \left[- (\alpha \frac{\partial
\beta}{\partial t} + \frac{\partial \nu}{\partial t})
 - {1 \over 2} (\alpha \vec \nabla \beta + \vec \nabla
\nu)^2 - \varepsilon(\rho) -\Phi \right] \rho d^{3}x.
\label{Ls}
\enq
The Lagrangian is quadratic in the spatial Clebsch
representation $\alpha \vec \nabla \beta + \vec \nabla \nu$,
and linear in the temporal "Clebsch representation":
$\alpha \frac{\partial\beta}{\partial t} +
\frac{\partial \nu}{\partial t}$.

\subsection{ Potentials as Lagrangian Variables}

The Lagrangian $L_s$ contains variables ($\alpha,\beta,\nu$)
that have no physical meaning outside the context
of the Clebsch representation. This is also the situation in
electromagnetics in which the Lagrangian
(see for example Goldstein (1980) p. 582) is given by:
\beq
L_{em}=\int_V \left[ \frac{1}{2} (
\frac{\partial \vec A }{\partial t}^2
+ 2 \vec \nabla A_0 \cdot \frac{\partial \vec A }{\partial t} +
\vec \nabla A_0^2 - \vec \nabla \times  \vec A^2) +
e (\vec j \cdot \vec A - \rho A_0) \right] d^{3} x.
\label{Lem}
\enq
In which $\vec A,A_0$ are the vector and scalar electromagnetic
potentials, $e$ is the charge of the electron, $\vec j$ is the
current and $\rho$ is the charge density\footnote{Not the
matter density!}.
Only after the problem is solved in terms of the potentials
$\vec A,A_0$ the physical electric $\vec E$ and magnetic $\vec
B$ fields can be obtained through the relations:
\beq
\vec E = -\frac{\partial \vec A }{\partial t} - \vec \nabla A_0
\qquad
\vec B = \vec \nabla \times \vec A
\enq
which are analogue to the Clebsch representation given by \er{clebsch}
in the fluid dynamical case.

\subsection{ Linearity in Time Derivatives }

Another odd characteristic of the Lagrangian $L_s$ given
by \er{Ls} is that it contains only linear terms in time
derivatives. This is unlike the generic case in classical
mechanics in which terms quadratic in time derivatives appear in the
kinetical energy part the Lagrangian.
For this situation I have been able to find an analogue from
quantum mechanics. The Schroedinger Lagrangian is given by:
\beq
L_{sch}= \int \left[ -i \hbar \frac{\partial \psi }{\partial t}
\psi^* + \frac{\hbar^2}{2m} \vec \nabla \psi \cdot \vec \nabla \psi^* +
V \psi \psi^*  \right] d^{3} x.
\label{Lsch}
\enq
in which $\hbar=\frac{h}{2 \pi}$ is the Planck Constant,
$i=\sqrt{-1}$, $\psi$ is the wave function and $\psi^*$ is
its complex conjugate.
$L_{sch}$ is linear in time derivatives and in this respect
resembles $L_s$.

\section{Conclusions}

Although a compact variational principle have been obtained in 
\er{Ls} it remains to see the possible implications of this expression.
One possible utility is to obtain better numerical algorithms for 
solving fluid flow problems based on the above variational principle 
(see Yahalom \cite{Yahalom}). Another may be to study the stability 
of certain stationary flows in the spirit of the works by Katz, Inagaki 
\& Yahalom \cite{Katz} and Yahalom, Katz \& Inagaki \cite{YahalomKatz}.

Further More, it is also desirable to have a variational principle for 
incompressible flows which should contain only three functions since the 
density is not a variable in this case. A similar reduction in degrees 
of freedom should be obtained for two dimensional flows in which the velocity 
has only two components. The above list of extensions of the 
Eulerian variational principles is of course not comprehensive. 
And the undertakings mentioned are left for future papers.

\end{document}